\documentclass[american,english,aps,preprint]{revtex4-1}
\usepackage[T1]{fontenc}
\usepackage[latin9]{inputenc}
\usepackage{units}
\usepackage{amsmath}
\usepackage{amssymb}
\usepackage{graphicx}
\usepackage{esint}
\usepackage{babel}
\begin{document}

\preprint{BROWN-HET-1678}

\title{Quantum information erasure inside black holes}

\author{David A. Lowe}

\email{lowe@brown.edu}

\affiliation{Department of Physics, Brown University, Providence, RI, 02912, USA}

\author{Larus Thorlacius}

\email{lth@hi.is}

\affiliation{University of Iceland, Science Institute, Dunhaga 3, IS-107 Reykjavik,
Iceland\foreignlanguage{american}{\\{\rm and}\\ }The Oskar Klein
Centre for Cosmoparticle Physics, Department of Physics, Stockholm
University, AlbaNova University Centre, 10691 Stockholm, Sweden}
\begin{abstract}
An effective field theory for infalling observers in the vicinity
of a quasi-static black hole is given in terms of a freely falling
lattice discretization. The lattice model successfully reproduces
the thermal spectrum of outgoing Hawking radiation, as was shown by
Corley and Jacobson, but can also be used to model observations made
by a typical low-energy observer who enters the black hole in free
fall at a prescribed time. The explicit short distance cutoff ensures
that, from the viewpoint of the infalling observer, any quantum information
that entered the black hole more than a scrambling time earlier has
been erased by the black hole singularity. This property, combined
with the requirement that outside observers need at least of order
the scrambling time to extract quantum information from the black
hole, ensures that a typical infalling observer does not encounter
drama upon crossing the black hole horizon in a theory where black
hole information is preserved for asymptotic observers. 
\end{abstract}
\maketitle

\section{introduction}

Recently the authors proposed a method to build quasi-local bulk operators
inside black hole horizons in unitary quantum theories \cite{Lowe:2014vfa}.
The method relies on two conjectured properties of such theories: 
\begin{enumerate}
\item that the minimal decoherence time of a black hole (or the time required
for outside observers to extract quantum information from the black
hole) has a lower bound of order the scrambling time, $t_{scr}=4M\log(4M),$
\footnote{More generally $t_{scr}=\mathcal{O}(\beta\log S),$ with $\beta$
the inverse Hawking temperature and $S$ the Bekenstein-Hawking entropy.}
\item that, from the viewpoint of an infalling observer who enters the black
hole, any quantum information that entered more than a scrambling
time earlier has been erased. 
\end{enumerate}
The construction in \cite{Lowe:2014vfa} is tailored to fit observations
made by an observer falling into a black hole at a prescribed time
and is only valid for a scrambling time before and after the observer
enters the horizon. During this restricted time period, the resulting
effective theory describes observations made by a typical low-energy
infalling observer, to within the limited experimental precision available
to such an observer, and its predictions agree with those the underlying
unitary theory to well within the observable precision.

In the present paper we consider a simple discretization of the radial
direction in an infalling reference frame and explore the consequences
for the construction of quasi-local bulk operators in the black hole
interior. This can be viewed as a concrete realization, with an explicit
short distance cutoff, of the effective field theory construction
in \cite{Lowe:2014vfa}. The same type of discretization has been
considered before in the work of Corley and Jacobson \cite{Corley:1997ef},
in a dimensionally reduced model for black hole evaporation. There
the focus was on the black hole exterior and showing that the spectrum
of the Hawking flux is insensitive to the discretization. It turns
out the model is also suitable for a bulk effective description of
low-energy observers who enter a black hole in free fall. As we shall
see below, the combination of large redshift factors in the black
hole region and having a strict short distance cutoff severely restricts
the amount of quantum information accessible to a typical infalling
observer inside the black hole. Combined with property \#1, this ensures
that a typical infalling observer does not encounter drama upon crossing
the black hole horizon \cite{Lowe:2014vfa}.

General covariance is the symmetry principle underlying Einstein's
theory of gravitation. The essence of the black hole information problem,
is that in a quantum theory, general covariance leads to a conflict
between unitarity and bulk locality. Holographic models, such as the
anti-de Sitter gravity/conformal field theory (AdS/CFT) correspondence,
provide a setting where the unitarity puzzle is resolved at the expense
of bulk locality. It is less well-understood, however, how the violation
of bulk locality avoids infecting observations made by low-energy
local observers both inside and outside the black hole in the limit
where one expects the semiclassical approximation for bulk gravity
to be good. We note that a finite $N$ reconstruction of the bulk
geometry from the holographic theory involves a soft violation of
general covariance. The symmetry is expected to be restored in the
$N\rightarrow\infty$ limit but Hawking emission is a $\nicefrac{1}{N}$
effect in holographic theories and the black hole information problem
cannot be studied in the strict $N\rightarrow\infty$ limit. In AdS/CFT
examples, symmetries associated with the transverse directions are
linearly realized in the holographic description, and such directions
are reconstructed exactly in the bulk theory. However, the bulk radial
direction arises from scale transformations, which are non-linearly
realized in the quantum holographic theory, and the breaking of general
covariance is implemented through a subtle discretization of the radial
direction. The details of this holographic regulator are currently
poorly understood, in particular when the bulk spacetime contains
a black hole. 

There is general consensus that quasi-local bulk operators can be
constructed in black hole exteriors on length scales approaching the
Planck length. In particular, the bulk effective Hamiltonian can be
constructed for a quantum field in the background of a large black
hole \cite{Kabat:2011rz}. One of the central ideas behind the construction
of \cite{Lowe:2014vfa}, was that evolution of exterior operators
under this Hamiltonian provides a natural definition of interior bulk
operators. The discretization we consider here provides a model where
the properties of such interior operators may be explored. The description
is only approximate and does not capture the holographic physics in
all detail. However, this is not a problem since observations made
by an inside observer in free fall necessarily have limited precision
both due to the restricted size of measuring devices that can be carried
into a black hole without back-reacting on the geometry and due to
the finite proper time remaining to the observer before encountering
the black hole singularity \cite{Lowe:2006xm}. Any effective field
theory or other model description for infalling observers need only
reproduce the exact holographic physics to within this limited precision
\cite{Lowe:2006xm}.

\section{Free falling lattice model}

In a black hole exterior, there seems to be considerable freedom in
defining bulk operators in holographic models. This makes sense, because
at least for arbitrarily large distances from the black hole, one
can consider effective field theories with large relative boosts with
respect to the black hole, without having to worry about issues of
back-reaction. Here by effective field theory, we mean a field theory
with a local Lagrangian, which may be accurately treated using a semiclassical
expansion, ignoring quantum gravitational effects.

Close to the black hole horizon, this is in general no longer the
case. One is not free to consider effective theory in an arbitrary
frame where typical energies are trans-Planckian with respect to the
black hole rest frame. Nevertheless, the equivalence principle dictates
that we should be able to construct an effective field theory in a
freely falling frame outside the black hole, where typical local frequencies
are much smaller than the Planck scale. From the holographic perspective,
this means we should be able to construct quasi-local operators, localized
on scales $\lambda$ in this frame satisfying $GM\gg\lambda\gg l_{Planck}$,
where $l_{Planck}$ is the Planck length.

Of course, one might argue that if the holographic theory is to truly
describe generally covariant gravity, it must also accommodate observers
or probes with arbitrarily large relative boosts with respect to the
black hole. At the end of the day, such a requirement imposes general
covariance on the quantum theory. It is clear, however, that the holographic
theory does not contain general covariance as a manifest symmetry
for finite $N$. Certainly in a large $N$ limit one hopes it is equivalent
to classical string theory in the background in question, but to see
the Hawking radiation in the first place, it is essential to keep
$N$ (or more generally the central charge of the conformal field
theory) finite. Our point of view will be that the holographic theory
is able to supply an effective field theory in a coordinate patch
satisfying the conditions of the previous paragraph, but is not fully
generally covariant, and our task is to study whether such an effective
field theory is sufficient to describe the physics of an infalling
observer as they cross the horizon, where general covariance does
emerge in a classical limit.

A two-dimensional free-falling lattice model was studied by Corley
and Jacobson \cite{Corley:1997ef}, where their main goal was to show
that the spectrum of Hawking radiation was insensitive to the details
of the short distance cutoff. In the present work, we use essentially
the same model, treating the radial direction as a free falling lattice
with Planck scale lattice spacing near the horizon. Rather than work
with a two-dimensional model, one can also generalize to four dimensions,
treating the angular directions as continuous spheres, but for our
purposes it turns out to be sufficient to consider the s-wave sector.
We will be focusing on near-horizon physics, so we restrict discussion
to the simple case of a Schwarzschild black hole in asymptotically
flat spacetime. Generalization to spacetimes with other asymptotics
will not affect our main conclusions. Furthermore, since we will be
considering timescales that are parametrically short compared to the
black hole lifetime, we will model the background geometry as a static
Schwarzschild spacetime.

The starting point for the lattice model is the Schwarzschild metric
in Gullstrand-Painlevé coordinates (in units with $G=\hbar=c=1)$.
We define
\begin{equation}
v(r)=-\sqrt{\frac{2M}{r}}\,,\label{eq:vmetric}
\end{equation}
and then the metric is
\[
ds^{2}=-\left(1-v^{2}(r)\right)dt^{2}-2v(r)dt\,dr+r^{2}d\Omega^{2}\,,
\]
where the time coordinate is obtained by adding an $r$-dependent
function to the usual Schwarzschild time
\[
t=t_{s}+2M\left(-\frac{2}{v(r)}-\log\left(\frac{1-v(r)}{1+v(r)}\right)\right)\,.
\]
The coordinate transformation is singular on the horizon, as it must
be for the Gullstrand-Painlevé coordinates to be smooth there. We
note, however, that this is traded for a coordinate singularity at
large $r$. 

For the moment, we take the metric to be of the form \eqref{eq:vmetric}
inside the horizon of the black hole $r<2M$. This amounts to assuming
that neither additional matter stress energy nor gravitational waves
are present inside the horizon. This may appear to be rather a strong
assumption given that the lattice model based on this metric is to
be viewed as a representation of the exact physics described by the
holographic theory. However, as we see later on (towards the end of
section \ref{sec:Dispersive-propagation}), there is a sense in which
the metric \eqref{eq:vmetric} behaves as an attractor in the regulated
theory.

The metric is stationary in Gullstrand-Painlevé coordinates so $p_{t}$
is conserved along a timelike geodesic. For a particle of unit mass
on a radial geodesic we have

\[
p_{t}=-E=-\left(1-v^{2}(r)\right)\frac{dt}{d\tau}-v(r)\frac{dr}{d\tau}\,.
\]
Imposing the usual normalization condition on the 4-velocity of the
particle, leads to the orbit equation for radial geodesics
\[
\left(\frac{dr}{d\tau}\right)^{2}=E^{2}-1+v^{2}(r)\,.
\]
The infall coordinates used in \cite{Corley:1997ef} are constructed
from geodesics with $E^{2}=1$, which are at rest at infinity. For
this choice, $t$ is equal to the proper time along the geodesic.
One may then introduce a new radial coordinate 
\[
y=t-\int_{2M}^{r}\frac{dr'}{v(r')}\,,
\]
which is constant along the geodesic. In these coordinates, the metric
is
\begin{equation}
ds^{2}=-dt^{2}+v^{2}(r)dy^{2}+r^{2}d\Omega^{2}\,,\label{eq:metric}
\end{equation}
with
\begin{equation}
r(y,t)=2M\left(1+\frac{3}{4M}\left(y-t\right)\right)^{2/3}\,.\label{eq:radius}
\end{equation}
The horizon is located at $y=t$ and the curvature singularity at
$y=t-\frac{4}{3}M$. In particular, the $y=0$ geodesic enters the
horizon at $t=0$, and hits the singularity at $t=\frac{4}{3}M$.

The freely falling lattice model is obtained by discretizing the $y$
coordinate. We choose a freely falling Planck scale lattice near the
horizon (rescaling $M$ can be used to rescale this to any desired
length), as motivated by holographic models. At larger radius, the
proper spacing falls below the Planck length, limiting the region
of spacetime where the effective field theory description will be
valid. However this will be sufficient for our purposes, and was already
sufficient to show cutoff independence of the Hawking flux. The breakdown
of the free-fall lattice regulator far from the black hole is in line
with black hole complementarity. Presumably to represent the far region
using a holographic description, one must evolve operators with respect
to a different time coordinate, such as the Schwarzschild time, resulting
in a very different regulator in the bulk effective field theory.

Let us consider a massless scalar field on the freely falling lattice.
We choose units such that the lattice spacing in $y$ is $1$. Since
the most dangerous modes for us are s-waves, it is convenient to truncate
to only those modes. The Lagrangian is then 
\begin{equation}
S=2\pi\sum_{y}\int dt\,r^{2}(y,t)\left(|v(r(y,t))|\left(\frac{\partial\phi}{\partial t}\right)^{2}-\frac{2\left(D_{y}\phi\right)^{2}}{|v(r(y+1,t))+v(r(y,t))|}\right)\,,\label{eq:action}
\end{equation}
with the difference operator
\[
D_{y}\phi=\phi(y+1,t)-\phi(y,t)\,.
\]
An attractive feature of this discretization is that the action has
a residue of the Killing symmetry of the Schwarzschild metric
\begin{equation}
(y,t)\to(y+1,t+1)\,.\label{eq:symmetry}
\end{equation}
In particular, this allows one to use separation of variables to derive
the form of the mode functions, and determine the lattice dispersion
relation. The general form of the mode function takes the form 
\begin{equation}
\phi(y,t)=e^{-i\omega t}f(y-t)\,,\label{eq:wavemode}
\end{equation}
where the frequency $\omega$ may be taken to be any real number.
The invariance of the mode function under
\begin{equation}
k\rightarrow k+2n\pi,\qquad\omega\rightarrow\omega-2n\pi\label{eq:shiftsymmetry}
\end{equation}
for any integer $n$ allows one to map an arbitrary pair $(\omega,k)$
into the range
\begin{eqnarray}
-\pi< & k & \leq\pi\nonumber \\
-\infty< & \omega & <\infty\,.\label{eq:moderange-1}
\end{eqnarray}

In a WKB approximation, we expand the mode function in terms of wavevectors
that depend on position
\begin{equation}
\phi(y,t)=e^{-i\omega t}e^{ik(r)(y-t)}\,.\label{eq:wkbmodes}
\end{equation}

Plugging \eqref{eq:wkbmodes} into the equation of motion, and assuming
$r(y,t)$ varies slowly on the length scale defined by the local wave
vector, yields the dispersion relation obtained by Corley and Jacobson
\cite{Corley:1997ef} for the WKB wavevector $k(r(y,t))$
\begin{equation}
|v(r)|(\omega+k)=\pm2\sin\left(k/2\right)\,.\label{eq:dispersion}
\end{equation}
\begin{figure}
\includegraphics{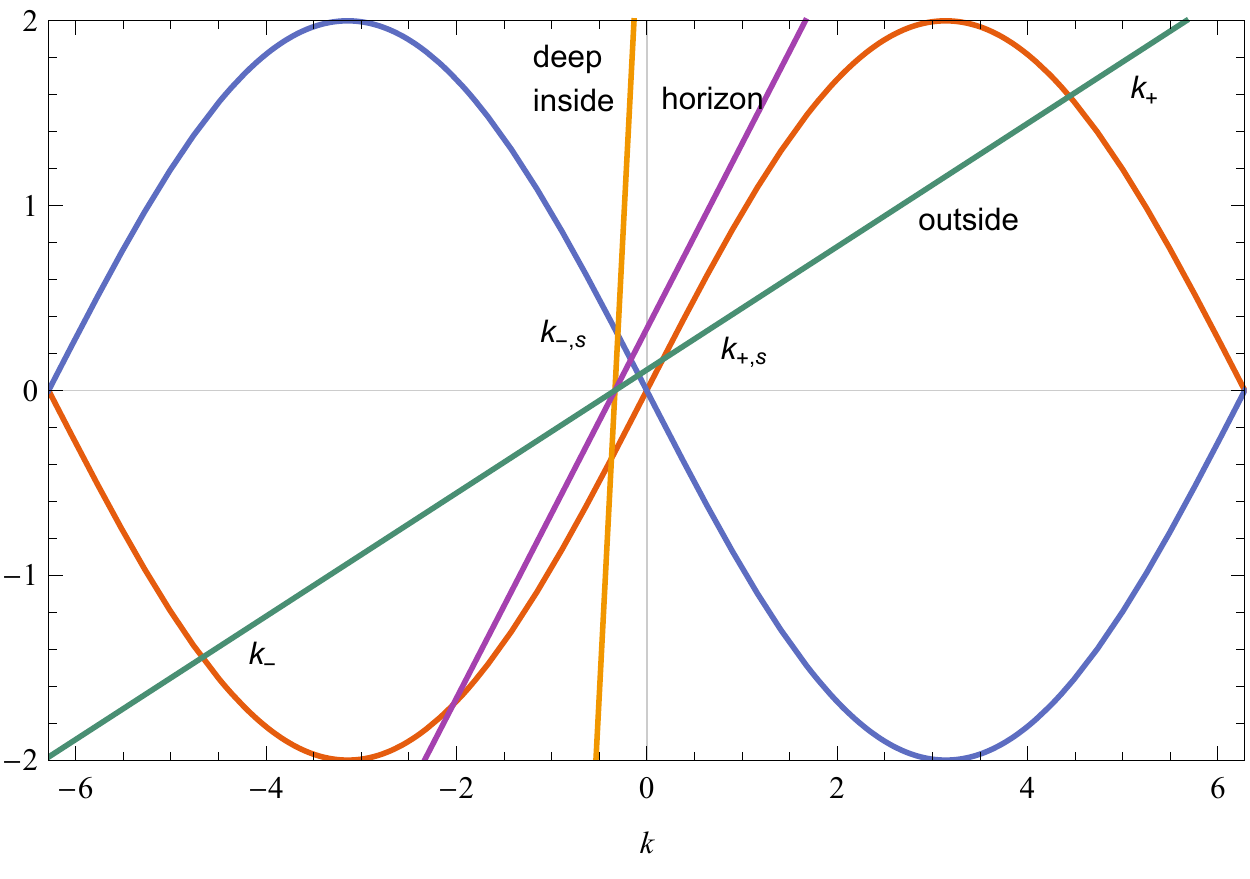}\caption{\label{fig:Dispersion-relation}Graphical solution of the lattice
dispersion relation. The straight lines correspond to the left hand
side of \eqref{eq:dispersion} for different values of $v(r)$ (and
$\omega>0$).}

\end{figure}
This dispersion relation governs the propagation of wavepackets in
the $(y,t)$ plane. It can be solved numerically but the qualitative
behavior of solutions can be obtained by simple graphical analysis
based on figure \ref{fig:Dispersion-relation}. The straight lines
in the figure correspond to the left-hand side of \eqref{eq:dispersion}
for $\omega>0$ and three different values of $r;$ outside, inside
and at the black hole horizon, with the slope of the line determined
by \eqref{eq:vmetric} in each case. Far away from the black hole
$v(r)\rightarrow0$ and the line approaches horizontal, while deep
inside the black hole it becomes vertical. For a given value of $r,$
the solutions to the dispersion relation are given by the points where
the corresponding straight line intersects the sinusoidal curves.
The sign of the group velocity in the $y$ coordinate
\[
\frac{dy}{dt}=\pm\frac{1}{|v(r)|}\cos\left(k/2\right)
\]
determines whether the mode is right- or left-moving. For each $\omega>0$
solution represented in figure \ref{fig:Dispersion-relation}, there
is a corresponding one with $\omega<0$ (and opposite sign $k$).
The two branches of solutions are related by complex conjugation and
we will restrict attention to the $\omega>0$ case. 

Far away from the black hole there are four intersection points, corresponding
to four different $\omega>0$ modes, of which two survive the continuum
limit and two are only present at finite lattice cutoff. Following
\cite{Corley:1997ef}, we denote the modes by $\psi_{+},$  $\psi_{-},$
$\psi_{+,s},$ and $\psi_{-,s},$ and let the corresponding WKB momenta
carry the same subscripts, as indicated in figure \ref{fig:Dispersion-relation}.
Of these modes, $\psi_{-,s},$ is left-moving while the other three
are right-moving. The momenta $k_{+}$ and $k_{-}$ lie outside the
fundamental domain in \eqref{eq:moderange-1} but can be brought inside
it by a shift of the form \eqref{eq:shiftsymmetry}. 

Free fall observers, traveling along lines of constant $y$ measure
the free-fall frequency
\[
\omega_{ff}=\omega+k\,,
\]
which is invariant under the shift symmetry \eqref{eq:shiftsymmetry}.
In particular, low-energy observers in free fall will only have access
to modes where $|\omega_{ff}|\ll1$. Note that a $\omega>0$ mode
can have negative free-fall frequency. This applies, for instance,
to the $k_{-}$ mode in figure \ref{fig:Dispersion-relation}.

As we move towards the black hole, the intersection points migrate
along the sinusoidal curves and at a critical value of $r$ the two
right-moving modes with positive $k$ merge, leading to a WKB turning
point outside the black hole \cite{Corley:1997ef}. At smaller values
of $r$ only two $\omega>0$ modes remain: $\psi_{-},$ which is right-moving,
and $\psi_{-,s},$ which is left-moving. 

It is worth commenting that the lattice gives rise to a continuum
of modes in $(\omega,k)$ space, which include wavepackets localized
behind the horizon. Presumably in the exact quantum theory there is
some further discretization leading to a finite number of states behind
the black hole horizon. It is reasonable this should be invisible
in any semiclassical approximation since the expected energy level
spacing will be of order $e^{-S_{bh}}$ where $S_{bh}$ is the Bekenstein-Hawking
entropy. Certainly such a refined level spacing is irrelevant to infalling
observers behind the horizon, who only live for a maximal proper time
of order the light-crossing time of the black hole.

\section{Dispersive propagation\label{sec:Dispersive-propagation}}

The bulk Hamiltonian in these coordinates generates evolution along
timelike geodesics from outside the horizon all the way to the singularity.
Unitarity is manifest, except for the component of the wavefunction
that hits the spacelike singularity. In effect, the collision of points
with the singularity replaces the curtains of fire suggested by some
authors\footnote{At the singularity one needs a prescription to evolve the state forward
in time, in order to preserve its normalization. A natural prescription
is to trace over modes that hit the singularity, mimicking how one
treats the exterior region in the original proposal of Hawking. } \cite{braunstein,Almheiri:2012rt}. The trick is to come up with
a self-consistent way to maintain the validity of the interior freely
falling description, while having a complementary unitary exterior
evolution. As discussed in detail in \cite{Lowe:2014vfa}, this can
be accomplished in theories of quantum gravity that obey properties
\#1 and \#2 listed in the Introduction. Our immediate goal is to derive
property \#2 within the lattice model of the previous section. This
provides a concrete example of the behavior argued for on general
grounds in our earlier work \cite{Lowe:2014vfa} and lends credence
to the self-consistency of our general effective field theory picture
for infalling observers put forward in that work.

We wish to estimate the correlation function of a wavepacket operator,
specified to be inside the horizon on a given time slice, with another
similarly constructed wavepacket evolved by free fall from outside
the horizon, crossing the horizon at a later time. The by now standard
theory of decoherence (for a textbook treatment see \cite{Weinberg:2012})
shows that local interactions with a large number of exterior degrees
of freedom decohere a quantum state, if given sufficient time. In
the case of a black hole this decoherence time is at least of order
the scrambling time \cite{Hayden:2007cs}. Correlation functions between
wavepackets that enter the horizon separated by less than a scrambling
time are not affected by decoherence. However, if our initial wavepacket
is capable of sending a detectable signal to a late time wavepacket
that enters the black hole a scrambling time or more later, then the
black hole complementarity principle will be falsified \cite{Susskind:1993mu}.
In other words, if one takes property \#1 as a postulate, then a violation
of property \#2 implies violation of black hole complementarity.

It is straightforward to analyze the propagation of such wavepackets,
using standard methods of propagation in dispersive media \cite{Brillouin-1960}.
The largest amplitude component of the wavepacket propagates within
a timelike cone, bounded by the evolution of trajectories according
to the group velocity of the different Fourier components. The group
velocity with respect to the $y-t$ coordinate of a wavetrain \eqref{eq:wavemode}
is
\begin{equation}
v_{g}=\frac{d\omega}{dk}=\pm\frac{\cos\left(k/2\right)}{|v|}-1\,.\label{eq:groupvelocity}
\end{equation}

\begin{figure}

\includegraphics[scale=0.6]{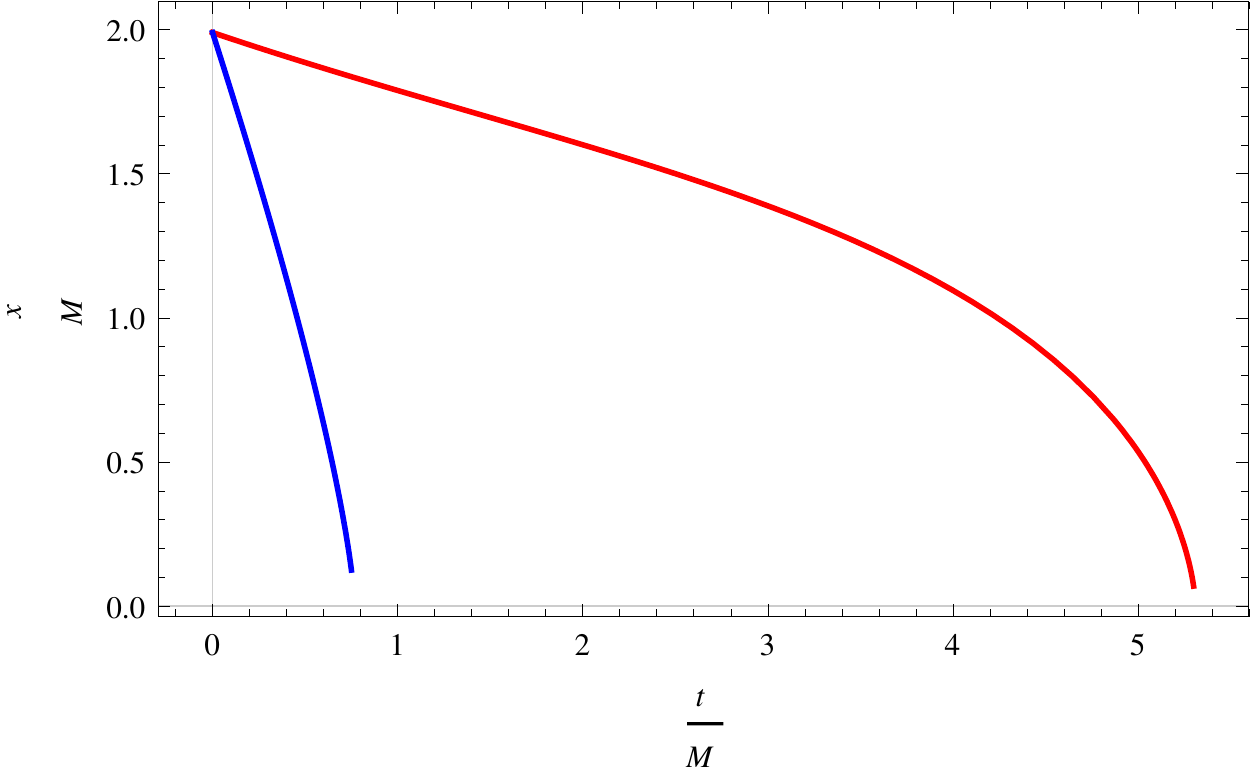}\includegraphics[scale=0.6]{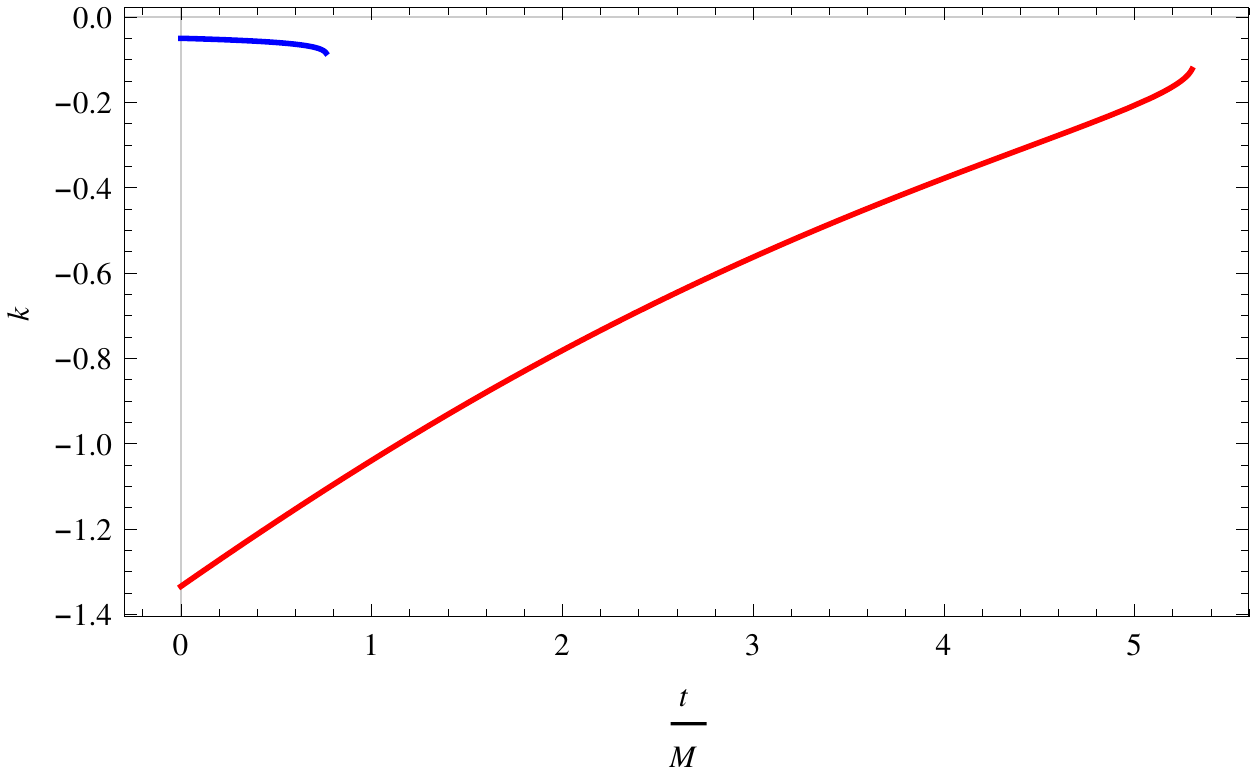}\caption{\label{fig:Interior-wavepacket-trajectories}Interior wavepacket trajectories.
The left figure shows the radial position as a function of the Gulstrand-Painlevé
coordinate $t$. The right figure shows the evolution of the WKB wavevector
$k$. The red curves represent the right-moving, horizon skimming
mode, which take a scrambling time to hit the singularity. The blue
curves represent the left-moving mode. In the figures, $M=100$, $\omega=1/10$.}
\end{figure}

Let us begin with a wavepacket just inside the horizon. This is inside
the future trapped region so both signs in \eqref{eq:groupvelocity}
correspond to ingoing wavepackets. The minus sign gives rise to a
left-moving wavepacket, i.e. one that moves towards negative $y$
in the infalling lattice, while the plus sign gives a right-moving
wavepacket, moving towards positive $y$. The plot on the left in
figure \ref{fig:Interior-wavepacket-trajectories} shows numerical
results for the radial position of left- and right-moving wavepackets
as a function of time, while the plot on the right tracks the value
of the wavevector $k_{max}$, where the wavepacket has maximum amplitude.
The left-moving mode rapidly hits the singularity, as expected, with
little change in $k_{max}$. The right-moving horizon-skimming modes
are of more interest to us as they take much longer to hit the singularity
and the WKB value for $k_{max}$ can evolve significantly along the
way. 

In figure \ref{fig:Interior-wavepacket-trajectories} the right-moving
wavepacket starts out near the horizon with a large wavevector so
the free fall frequency $\omega_{ff}$ of this mode is large compared
to $\omega$. In view of this, it is instructive to compare the free
fall frequencies of left- and right-moving wavepackets near the horizon
in the limit $\omega\ll1$. Inserting $v=-1$ into \eqref{eq:dispersion}
and solving for small $|k|$ gives $\omega_{ff}\approx-k$ for left-moving
modes, which is of the same order of magnitude as $\omega$. For a
right-moving wavepacket one instead finds 
\begin{equation}
\omega_{ff}\approx(24\omega)^{1/3}\,.\label{eq:bigk}
\end{equation}
The appearance of this intermediate scale in the regulated theory
was noted in \cite{Corley:1997ef}, though in that case the focus
was on exterior modes, where the same $\omega^{1/3}$ behavior is
seen near the classical turning point in the WKB approximation. The
existence of such an intermediate length scale in the regulated theory
may be of interest in the context of the non-local information transfer
across the event horizon proposed in \cite{Giddings:2013kcj}.

Inside the horizon, as is clear from figure \ref{fig:Dispersion-relation},
there is no classical turning point, so a straightforward WKB approximation
should be valid for interior wavepackets until they approach the curvature
singularity. Let us consider the correlator of a right-moving interior
$\psi_{-}$ wavepacket and a later left-moving $\psi_{s,-}$ wavepacket,
separated by a time $\delta t\approx t_{scr}$. First we estimate
the width of the $\psi_{-}$ wavepacket, by the time it has propagated
to the late time slice. To proceed, let us begin with a wavepacket
of the form
\begin{equation}
\psi_{initial}(y,t)=\int d\omega\,e^{-(\omega-\omega_{0})^{2}a^{2}}e^{ik_{-}(\omega,r)(y-t+d)-i\omega t}\,,\label{eq:initialpacket}
\end{equation}
where $d\gg a>1$ and $\omega_{0}\sim1/a$ in order that the wavepacket
be well-localized at $y=-d$ inside the horizon at $t=0$. The dominant
components of the wavepacket travel at the group velocity \eqref{eq:groupvelocity}
(with + sign) which is less than the speed of light with respect to
the infalling lattice. The time it takes the dominant part of the
wavepacket to hit the singularity is therefore bounded from above
by the time it takes a right-moving radial null geodesic to hit the
singularity starting from the an initial point $(y,t)=(-1,0)$ just
inside the horizon. On such a geodesic
\[
\frac{dy}{dt}=-\frac{1}{v(r)}\,.
\]
Using \eqref{eq:radius} one obtains
\[
\frac{dr}{dt}=1+v(r)\,,
\]
which integrates to 
\[
\frac{t}{4M}=\frac{r}{4M}-\frac{1}{v(r)}+\log\left|1+\frac{1}{v(r)}\right|+A\,,
\]
where $A$ is an integration constant to be determined by the condition
it pass through the initial point. For large $M$
\[
A\approx\log\left(4M\right)-\frac{3}{2}+\mathcal{O}(1/M)\,.
\]
The geodesic hits the singularity at 
\[
t=4MA\approx4M\log4M\,,
\]
which we define as the scrambling time $t_{scr}.$ 

Returning to our initial right-moving wavepacket, we see from this
that it will hit the singularity no later than at the scrambling time.
As the singularity is approached the metric diverges, $|v(r)|\rightarrow\infty$,
in which case the dispersion relation \eqref{eq:dispersion} reduces
to $\omega=-k$ and the width in $y$ of the wavepacket becomes of
order $a$. After a scrambling time the singlularity is located at
$y\approx t_{scr}$, and it follows that the overlap of the initial
wavepacket with another incoming wavepacket at the horizon is at most
of order $e^{-t_{scr}^{2}/a^{2}}$ in the WKB approximation. 

In addition to the dominant components of the wavepacket that travel
at the group velocity, there are the so-called forerunners \cite{Brillouin-1960}.
These are modes with different frequencies than the dominant component
of the signal and can travel faster than the dominant group velocity.
Such modes are substantially excited only when the group velocity
has a maximum or minimum, which is when the straight lines in figure
\ref{fig:Dispersion-relation} are tangent to the sine curves. The
only such turning point for the modes appears outside the horizon,
so this is not relevant for the interior propagation. In the absence
of a WKB turning point, the forerunners of an interior wavepacket
form a tail of low-frequency modes with $k_{-}\approx-(24\omega)^{1/3}$.
This tail essentially corresponds to the part of the initial wavepacket
that is outside the horizon at the initial time and its size can be
estimated by evaluating the integral in \eqref{eq:initialpacket}
in a saddle point approximation. The resulting contribution to the
overlap with the late time wavepacket is of order $e^{-c\,d^{2}/a^{2/3}}$
, where $c$ is an order one numerical constant, and is strongly suppressed
when $d\gg a>1$.

Thus far we have simply assumed an interior vacuum spacetime using
the metric \eqref{eq:vmetric} inside the horizon. However we have
concluded that both massive and massless matter excitations inside
the horizon hit the singularity within a scrambling time in the lattice
model. The same will be true for metric perturbations treated as background
fields in a lattice model and we conclude that the interior metric
effectively relaxes back to the vacuum within a scrambling time. Therefore
from the viewpoint of the holographic theory, assuming the metric
\eqref{eq:vmetric} should be reasonable as long as the amount of
matter that enters the black hole during a scrambling time prior is
insufficient to cause significant back-reaction on the geometry. 

As a potential counterexample, consider a right-moving negative energy
shockwave crossing the path of an infalling observer. This can cause
the $M$ parameter of \eqref{eq:vmetric} to decrease by an arbitrarily
large amount (after performing a suitable change of coordinates).
This would invalidate the conclusions of the previous section, where
it was argued the dispersion relation leads to information erasure
within a scrambling time. This is an extreme example of the so-called
``pots of gold'' solutions obtained by patching across domain walls
of stress energy \cite{Farhi:1989yr,Alberghi:1999kd,Lowe:2010np}.
If one imposes reasonable energy conditions on the domain walls, however,
these dangerous classical solutions are removed.

One might still worry about the effect of back-reaction of the Hawking
emission itself on the interior metric. However a number of Hawking
particles of order $\log S$ are emitted during a scrambling time,
and their typical energies are of order the Hawking temperature. Therefore
we do not expect a large effect on the interior metric since our patch
is only to be used for a range of times of order the scrambling time.

A detectable late-time signal in the interior would require an enormous
amplitude, either for the detector or the emitter, or both. Due to
the enormous exponential suppression factor, in either case, the required
trans-Planckian energy of the detector or emitter would produce enormous
back-reaction on the black hole. We conclude therefore, that for low-energy
observers, quantum information is effectively erased in the black
hole interior after a scrambling time has passed. This establishes
the validity of property \#2 in this model.

\section{Number operators}

One immediate cause for concern is that virtual modes with relatively
large free fall frequencies such as \eqref{eq:bigk} will cause apparent
violations of general covariance for a freely falling timelike observer,
due to their large free fall frequency, despite having a small frequency
$\omega$ with respect to the timelike Killing vector. 

Outside the horizon, there are four solutions to the dispersion relation:
$\psi_{+}\,,\psi_{-}\,,\psi_{+,s}$ and $\psi_{-,s}$ (in fact even
more solutions may appear at very large $r$). The corresponding wavevectors
are indicated in figure \ref{fig:Dispersion-relation}. A central
result of \cite{Corley:1997ef} is that the low-energy late-time outgoing
mode $\psi_{+,s}$ is entangled with the modes $\psi_{-}$ and $\psi_{+}$
if one imposes a free fall vacuum state far from the horizon. In particular,
the occupation number of the late-time $\psi_{+,s}$ modes is simply
a Bose-Einstein distribution with respect to the frequency $\omega$,
thus the lattice model reproduces the expected Hawking spectrum
\begin{equation}
N_{\omega}=\frac{1}{e^{\omega/kT}-1}\label{eq:boltzmann}
\end{equation}
where $T$ is the Hawking temperature of the black hole.

For the present work, it is also interesting to consider the number
operators associated with other wavepackets of interest. Some formalism
helpful in setting up these questions appears in \cite{Corley:1996ar}.
The easiest modes to consider are the $\psi_{-,s}$ modes. These correspond
to ingoing modes that freely propagate all the way to the singularity.
Because the WKB approximation is valid all the way along their trajectory,
there is negligible scattering into other modes. Hence these modes
have no effect on the late-time outgoing mode $\psi_{+,s}$. Their
free fall frequency is positive, so they are annihilated by the initial
free fall vacuum. The number operator $N_{-,s}$ will therefore have
vanishing expectation value in the free fall vacuum.

Now consider the large wavevector mode $\psi_{+}$. This mode has
positive free fall frequency, so at early times is annihilated by
the free fall vacuum. At late times, the mode propagates to a classical
turning point outside the horizon, where it undergoes mode conversion
to the outgoing mode $\psi_{+,s}$ and also tunnels into the interior.
Wavepackets corresponding to this mode can be constructed prior to
the mode hitting the classical turning point. Clearly evolving these
back in time, the WKB approximation remains valid, and the associated
number operator will have vanishing expectation value. Wavepackets
after the turning point correspond to the outgoing mode $\psi_{+,s}$
that we discussed above.

Finally, let us consider the large wavevector mode $\psi_{-}$. This
has negative free fall frequency, so the annihilation operator corresponds
instead to the conjugate mode $\psi_{-}^{*}$. It is possible to build
wavepackets dominated by the $\psi_{-}^{*}$ mode. In the continuum
limit, these modes would be localized inside the horizon (as is easily
seen by replacing the sine curves in \ref{fig:Dispersion-relation}
by straight lines). However with the regulator, the modes also have
support outside the horizon. This is a crucial new feature of the
regulated theory. Building a wavepacket well-approximated by such
a mode will evolve across the horizon and continue toward the singularity,
again with the WKB approximation remaining valid. We conclude that
the number operator of such a wavepacket will have vanishing expectation
value, since it can be smoothly propagated back to the initial timeslice
where the $\psi_{-}^{*}$ mode annihilates the free fall vacuum.

It is worth reiterating that while a Fock space built using the WKB
modes appears larger than the Fock space of the continuum theory,
one is not adding extra degrees of freedom. The action for the fields
\eqref{eq:action} remains that of a local scalar field with a lattice
regulator. 

It is interesting to compare these results with the conclusions of
\cite{Marolf:2013dba}, who argued that if the infalling number operators
were built out of exterior operators, a firewall would result. An
alternative proposal for building interior operators for old black
holes was proposed in \cite{Avery:2013vwa,Avery:2015hia}, by embedding
the exact holographic state in a larger semiclassical Hilbert space
using ancillae modes set in their ground state. In the lattice model
presently considered we see the ancillae directly arise in the exterior
region in the form of the WKB modes $\psi_{+}$ and $\psi_{-}$, providing
a concrete realization of the ideas of \cite{Avery:2015hia}. If the
lattice model correctly captures the relevant physics of the holographic
theory, then we see that ancillae modes arise, and that their initial
state is determined by the exterior state of the black hole via their
free fall vacuum state. In the continuum limit, the mode $\psi_{+}$
is never present, and the mode $\psi_{-}$ has support only inside
the black hole horizon, rather than at arbitrarily large distance
from the horizon as in the lattice model. 

The $\psi_{-}$ is the dangerous mode from the point of view of the
information problem. If not correctly entangled with the outgoing
mode $\psi_{+,s}$ via the free fall vacuum condition, the expectation
value of the stress energy tensor will be large. We note in the regulated
theory, this is even true well outside the horizon. In the continuum
limit the issue appears only across the horizon leading to the so-called
firewall proposal \cite{Almheiri:2012rt,braunstein}. 

The regulated effective description described in the present work
is only valid inside and near the horizon, and is suitable only for
the approximate representation of observations by an infalling observer,
who necessarily has a finite lifetime, and corresponding finite experimental
uncertainties. The regulated effective description gives rise to Hawking
radiation with the usual thermal spectrum \cite{Corley:1997ef}, but
it imparts no information about the black hole state Hawking radiation.
As emphasized in \cite{Lowe:2014vfa}, this restricts its usefulness
to a limited range in time extending a scrambling time before and
after the infalling observer enters the black hole, but this range
is sufficient to correctly describe observations made by the infalling
observer to within the available experimental precision. 

A complementary effective field theory description will be valid for
purely exterior observers whose observations must coincide with the
exact unitary description of the black hole evolution, but this requires
cutting out a timelike stretched horizon, where the boundary conditions
can no longer be represented by an ordinary local effective field
theory. It also requires using a regulator that smoothly extends to
infinity and is naturally formulated on the usual Schwarzschild coordinate
patch rather than the infalling Gullstrand-Painlevé coordinates which
break down in the asymptotic region.

If on the other hand, we had made the inconsistent assumption, that
is the foundation of the firewall argument, that a single effective
field theory description allows one to describe the black hole interior
and the exact outgoing Hawking radiation, then the $\psi_{-}$ mode
would necessarily not be entangled with $\psi_{+,s}$. In that case,
there would be a firewall inside the horizon in the continuum theory
and introducing a lattice regulator would lead to the firewall spreading
outside the horizon into the exterior region, since information retrieval
would be incompatible with the initial free fall vacuum.

\section{Equivalence principle violations}

One of the main conclusions of the present work and that of \cite{Corley:1997ef}
is that it is very difficult to construct experiments that detect
violations of general covariance. Hawking radiation appears to emerge
with a largely undistorted spectrum, and experiments in the vicinity
of black holes seem compatible with general covariance, as long as
Planck scale accelerations are not invoked.

It is of interest then to see if the lattice model, which arguably
represents the behavior of a wide class of holographic theories, offers
any distinctive signals of the violation of general covariance. The
simplest experiment to construct simply involves scattering a wavepacket
far from the black hole, off the horizon, to be observed later. This
involves the same mode function computations as in \cite{Corley:1997ef},
the difference being we now consider a real quantum wavepacket scattering,
rather than a virtual one. Tracing back the late time signal through
the turning point, and back out into the far region, we obtain a mode
composed of the two WKB modes $\psi_{+}$ and $\psi_{-}$ corresponding
to the large $|k|$ solutions of the dispersion relation \eqref{eq:dispersion}
at large $r$. There will also be an unscattered component that we
denote $\psi_{early}$ corresponding to the small $|k|$ solution
of the dispersion relation. The creation operator for the late-time
mode will take the form
\[
a_{late}^{\dagger}=a_{+}^{\dagger}+a_{-}
\]
when evolved back to the initial timeslice. The free-fall initial
state is annihilated by the $a_{-}$ operator. The interesting component
is then the piece corresponding to $a_{+}^{\dagger}$. 

This component of the wavepacket has the interesting property that
it scatters directly off the black hole horizon, despite being an
infalling mode. Ordinarily, the low $k$ mode with $k<0$ propagates
to the singularity with negligible scattering off the geometry. Instead,
we find if this large $k$ ingoing mode is excited, it propagates
to a turning point outside the horizon, and does a combination of
back-scattering out, and tunneling inside. This is a dramatic new
feature of horizon scattering, that conceivably could give rise to
detectable signatures. 

To better understand this mode, we convert from the coordinate large
$k$ to the proper wavevector in the far region. To do this we must
shift $k\to k-2\pi$, $\omega\to\omega+2\pi$ according to the symmetry
\eqref{eq:symmetry} and use $k_{proper}=k/|v(r)|$. In the low $\omega$
limit, this gives $k_{proper}=-2\pi-\omega$ in units of the cutoff,
and the proper frequency becomes $\omega+2\pi$. If the cutoff is
taken close to the Planck scale, this indicates $\psi_{+}$ corresponds
to a Planck energy mode.

Nature appears to naturally produce jets of highly energetic particles
in the vicinity of rotating black holes. The model suggests that such
particles can scatter directly off the horizon with amplitude of order
1. To see this, note the computation is
\[
\left\langle 0|a_{late}a_{late}^{\dagger}|0\right\rangle =1+\frac{1}{e^{\omega/kT}-1}
\]
using $aa^{\dagger}-a^{\dagger}a=1$ and \eqref{eq:boltzmann}, with
$|0\rangle$ the initial infalling vacuum state.

\section{Conclusions}

Without a cutoff, a massless field in a black hole background has
modes with very low frequency, when measured via the timelike Killing
vector, but very short wavelength, that potentially can carry information
about the black hole initial state to late time interior observers.
This process is incompatible with the black hole complementarity resolution
of the information paradox.

In this paper we have explored a lattice model containing features
expected in an effective description derived from a more fundamental
holographic theory, such as AdS/CFT. The cutoff eliminates these dangerous
modes, and introduces a number of new features which we have explored.
We conclude that quantum information that enters the black hole horizon
is erased in a scrambling time from the viewpoint of interior observers.
This supports one of the key conjectures of the proposal of \cite{Lowe:2014vfa}
for building consistent interior observables within a framework that
solves the information puzzle.
\begin{acknowledgments}
The research of D.L. was supported in part by DOE grant DE-SC0010010
and an FQXi grant. The research of L.T. was supported in part by Icelandic
Research Fund grant 130131-053, the University of Iceland Research
Fund, and the Swedish Research Council under contract 621-2014-5838.
\end{acknowledgments}

\bibliographystyle{utphys}
\bibliography{firewall3}

\providecommand{\href}[2]{#2}\begingroup\raggedright\begin{thebibliography}{10}

\bibitem{Lowe:2014vfa}
D.~A. Lowe and L.~Thorlacius, ``{Black hole complementarity: The inside
  view},'' \href{http://dx.doi.org/10.1016/j.physletb.2014.08.062}{{\em
  Phys.Lett.} {\bfseries B737} (2014) 320--324},
\href{http://arxiv.org/abs/1402.4545}{{\ttfamily arXiv:1402.4545 [hep-th]}}.
%%CITATION = ARXIV:1402.4545;%%.

\bibitem{Note1}
More generally $t_{scr}=\protect \mathcal {O}(\beta \protect \qopname \relax
  o{log}S),$ with $\beta $ the inverse Hawking temperature and $S$ the
  Bekenstein-Hawking entropy.

\bibitem{Corley:1997ef}
S.~Corley and T.~Jacobson, ``{Lattice black holes},''
  \href{http://dx.doi.org/10.1103/PhysRevD.57.6269}{{\em Phys.Rev.} {\bfseries
  D57} (1998) 6269--6279},
\href{http://arxiv.org/abs/hep-th/9709166}{{\ttfamily arXiv:hep-th/9709166
  [hep-th]}}.
%%CITATION = HEP-TH/9709166;%%.

\bibitem{Kabat:2011rz}
D.~Kabat, G.~Lifschytz, and D.~A. Lowe, ``{Constructing local bulk observables
  in interacting AdS/CFT},''
  \href{http://dx.doi.org/10.1103/PhysRevD.83.106009}{{\em Phys.Rev.}
  {\bfseries D83} (2011) 106009},
\href{http://arxiv.org/abs/1102.2910}{{\ttfamily arXiv:1102.2910 [hep-th]}}.
%%CITATION = ARXIV:1102.2910;%%.

\bibitem{Lowe:2006xm}
D.~A. Lowe and L.~Thorlacius, ``{Remarks on the black hole information
  problem},'' \href{http://dx.doi.org/10.1103/PhysRevD.73.104027}{{\em
  Phys.Rev.} {\bfseries D73} (2006) 104027},
\href{http://arxiv.org/abs/hep-th/0601059}{{\ttfamily arXiv:hep-th/0601059
  [hep-th]}}.
%%CITATION = HEP-TH/0601059;%%.

\bibitem{Note2}
At the singularity one needs a prescription to evolve the state forward in
  time, in order to preserve its normalization. A natural prescription is to
  trace over modes that hit the singularity, mimicking how one treats the
  exterior region in the original proposal of Hawking.

\bibitem{braunstein}
S.~L. Braunstein, S.~Pirandola, and K.~\ifmmode~\dot{Z}\else
  \.{Z}\fi{}yczkowski, ``Better late than never: Information retrieval from
  black holes,'' \href{http://dx.doi.org/10.1103/PhysRevLett.110.101301}{{\em
  Phys. Rev. Lett.} {\bfseries 110} (Mar, 2013) 101301}.
  \url{http://link.aps.org/doi/10.1103/PhysRevLett.110.101301}.

\bibitem{Almheiri:2012rt}
A.~Almheiri, D.~Marolf, J.~Polchinski, and J.~Sully, ``{Black Holes:
  Complementarity or Firewalls?},''
  \href{http://dx.doi.org/10.1007/JHEP02(2013)062}{{\em JHEP} {\bfseries 1302}
  (2013) 062},
\href{http://arxiv.org/abs/1207.3123}{{\ttfamily arXiv:1207.3123 [hep-th]}}.
%%CITATION = ARXIV:1207.3123;%%.

\bibitem{Weinberg:2012}
S.~Weinberg, {\em Lectures on quantum mechanics}.
\newblock Cambridge University Press, 2012.

\bibitem{Hayden:2007cs}
P.~Hayden and J.~Preskill, ``{Black holes as mirrors: Quantum information in
  random subsystems},''
  \href{http://dx.doi.org/10.1088/1126-6708/2007/09/120}{{\em JHEP} {\bfseries
  0709} (2007) 120},
\href{http://arxiv.org/abs/0708.4025}{{\ttfamily arXiv:0708.4025 [hep-th]}}.
%%CITATION = ARXIV:0708.4025;%%.

\bibitem{Susskind:1993mu}
L.~Susskind and L.~Thorlacius, ``{Gedanken experiments involving black
  holes},'' \href{http://dx.doi.org/10.1103/PhysRevD.49.966}{{\em Phys.Rev.}
  {\bfseries D49} (1994) 966--974},
\href{http://arxiv.org/abs/hep-th/9308100}{{\ttfamily arXiv:hep-th/9308100
  [hep-th]}}.
%%CITATION = HEP-TH/9308100;%%.

\bibitem{Brillouin-1960}
L.~Brillouin, {\em Wave propagation and group velocity}.
\newblock Academic Press, 1960.

\bibitem{Giddings:2013kcj}
S.~B. Giddings, ``{Nonviolent information transfer from black holes: a field
  theory parameterization},''
  \href{http://dx.doi.org/10.1103/PhysRevD.88.024018}{{\em Phys.Rev.}
  {\bfseries D88} (2013) 024018},
\href{http://arxiv.org/abs/1302.2613}{{\ttfamily arXiv:1302.2613 [hep-th]}}.
%%CITATION = ARXIV:1302.2613;%%.

\bibitem{Farhi:1989yr}
E.~Farhi, A.~H. Guth, and J.~Guven, ``{Is It Possible to Create a Universe in
  the Laboratory by Quantum Tunneling?},''
\href{http://dx.doi.org/10.1016/0550-3213(90)90357-J}{{\em Nucl. Phys.}
  {\bfseries B339} (1990) 417--490}.
%%CITATION = NUPHA,B339,417;%%.

\bibitem{Alberghi:1999kd}
G.~L. Alberghi, D.~A. Lowe, and M.~Trodden, ``{Charged false vacuum bubbles and
  the AdS / CFT correspondence},''
  \href{http://dx.doi.org/10.1088/1126-6708/1999/07/020}{{\em JHEP} {\bfseries
  07} (1999) 020},
\href{http://arxiv.org/abs/hep-th/9906047}{{\ttfamily arXiv:hep-th/9906047
  [hep-th]}}.
%%CITATION = HEP-TH/9906047;%%.

\bibitem{Lowe:2010np}
D.~A. Lowe and S.~Roy, ``{Punctuated eternal inflation via AdS/CFT},''
  \href{http://dx.doi.org/10.1103/PhysRevD.82.063508}{{\em Phys. Rev.}
  {\bfseries D82} (2010) 063508},
\href{http://arxiv.org/abs/1004.1402}{{\ttfamily arXiv:1004.1402 [hep-th]}}.
%%CITATION = ARXIV:1004.1402;%%.

\bibitem{Corley:1996ar}
S.~Corley and T.~Jacobson, ``{Hawking spectrum and high frequency
  dispersion},'' \href{http://dx.doi.org/10.1103/PhysRevD.54.1568}{{\em Phys.
  Rev.} {\bfseries D54} (1996) 1568--1586},
\href{http://arxiv.org/abs/hep-th/9601073}{{\ttfamily arXiv:hep-th/9601073
  [hep-th]}}.
%%CITATION = HEP-TH/9601073;%%.

\bibitem{Marolf:2013dba}
D.~Marolf and J.~Polchinski, ``{Gauge/Gravity Duality and the Black Hole
  Interior},'' \href{http://dx.doi.org/10.1103/PhysRevLett.111.171301}{{\em
  Phys.Rev.Lett.} {\bfseries 111} (2013) 171301},
\href{http://arxiv.org/abs/1307.4706}{{\ttfamily arXiv:1307.4706 [hep-th]}}.
%%CITATION = ARXIV:1307.4706;%%.

\bibitem{Avery:2013vwa}
S.~G. Avery and D.~A. Lowe, ``{Event horizons and holography},''
\href{http://arxiv.org/abs/1310.7999}{{\ttfamily arXiv:1310.7999 [hep-th]}}.
%%CITATION = ARXIV:1310.7999;%%.

\bibitem{Avery:2015hia}
S.~G. Avery and D.~A. Lowe, ``{Typical Event Horizons in AdS/CFT},''
\href{http://arxiv.org/abs/1501.05573}{{\ttfamily arXiv:1501.05573 [hep-th]}}.
%%CITATION = ARXIV:1501.05573;%%.

\end{thebibliography}\endgroup

\end{document}